\documentclass[aps, prd,twocolumn, amsmath, floats,floatfix,
superscriptaddress, nofootinbib,showpacs]{revtex4}

\usepackage{stackengine} 
\usepackage{doi}
\usepackage{array} \usepackage{ifpdf}
\ifpdf \usepackage[pdftex]{graphicx} \graphicspath{{figures/pdf/}}
\DeclareGraphicsExtensions{.pdf} \else \usepackage[dvips]{graphicx}
\usepackage{graphicx} \usepackage{pstricks} \usepackage{pst-node}
\usepackage{pst-blur} \graphicspath{{figures/eps/}}
\DeclareGraphicsExtensions{.eps} \definecolor{Pink}{rgb}{1.,0.75,0.8}
\definecolor{cyan2}{cmyk}{0.40,0,0,0} \fi
\usepackage{tikz}
\usetikzlibrary{positioning}
\usetikzlibrary{arrows}

\usepackage{float}

\usepackage{csquotes}

\usepackage{tabularx}
\usepackage{array}
\newcolumntype{Z}{>{\centering\let\newline\\\arraybackslash\hspace{0pt}}X}

\usepackage{bbm}
\usepackage{xspace}

\def\ET{{\sc Einstein Toolkit}\xspace}
\def\SENR{{\sc SENR/NRPy+}\xspace}

\def\p0{\partial_0}

\newcommand\IncG[2][]{\addstackgap{\raisebox{-.1\height}{\includegraphics[#1]{#2}}}}

\def\apj{Astrophys. J.}  \def\aap{Astron. Astrophys.}
\def\mnras{Mon. Not. R. Astron. Soc.}  \def\prd{Phys. Rev. D}
  
\def\apjl{Astrophys. J. Letters}

\usepackage{tikz}
\usepackage{amsmath} \usepackage{amssymb} \usepackage{listings}
\usepackage{mathtools} \usepackage{subfigure}

\begin{document}

\title{Numerical relativity in spherical coordinates with the Einstein Toolkit}

\author{Vassilios Mewes} \affiliation{Center for Computational Relativity and Gravitation, 
Rochester Institute of Technology, Rochester, NY 14623}
\author{Yosef Zlochower} \affiliation{Center for Computational Relativity and Gravitation, 
Rochester Institute of Technology, Rochester, NY 14623}
\author{Manuela Campanelli} \affiliation{Center for Computational Relativity and Gravitation, 
Rochester Institute of Technology, Rochester, NY 14623}
\author{\\Ian Ruchlin} \affiliation{Department of Mathematics, West Virginia University, 
Morgantown, West Virginia 26506, USA}
\author{Zachariah B. Etienne} \affiliation{Department of Mathematics, West Virginia University, 
Morgantown, West Virginia 26506, USA} \affiliation{Center for Gravitational Waves and 
Cosmology, West Virginia University, Chestnut Ridge Research Building, Morgantown, 
West Virginia 26505, USA}
\author{Thomas W. Baumgarte} \affiliation{Department of Physics and Astronomy, 
Bowdoin College, Brunswick, Maine 04011, USA}

\begin{abstract}
Numerical relativity codes that do not make assumptions on spatial symmetries most 
commonly adopt Cartesian coordinates.  While these coordinates have many attractive 
features, spherical coordinates are much better suited to take advantage of approximate 
symmetries in a number of astrophysical objects, including single stars, black holes 
and accretion disks.  While the appearance of coordinate singularities often spoils 
numerical relativity simulations in spherical coordinates, especially in the absence 
of any symmetry assumptions, it has recently been demonstrated that these problems 
can be avoided if the coordinate singularities are handled analytically.  This is 
possible with the help of a reference-metric version of the Baumgarte-Shapiro-Shibata-Nakamura 
formulation together with a proper rescaling of tensorial quantities.  In this paper we 
report on an implementation of this formalism in the \ET.  We adapt the \ET 
infrastructure, originally designed for Cartesian coordinates, to handle spherical 
coordinates, by providing appropriate boundary conditions at both inner and outer 
boundaries.  We perform numerical simulations for a disturbed Kerr black hole, 
extract the gravitational wave signal, and demonstrate that the noise in these 
signals is orders of magnitude smaller when computed on spherical grids rather 
than Cartesian grids.  With the public release of our new \ET thorns, our methods 
for numerical relativity in spherical coordinates will become available to the 
entire numerical relativity community.
\end{abstract}


\maketitle

\section{Introduction}
\label{sec:introduction}
LIGO~\cite{Abramovici1992,Harry2010} and Virgo's~\cite{Bradaschia1990,Accadia2012}
first direct detections of gravitational waves (GWs) 
from both binary black hole (BBH) and binary neutron star (BNS) 
mergers~\cite{Abbott2016_GW150914,Abbott2016_GW151226,Abbott2017_GW170104,
Abbott2017_GW170608,Abbott2017_GW170814,Abbott2017_GW170817} open 
a new window for observations of the Universe. Moreover, the simultaneous detection 
of GWs and electromagnetic (EM) radiation from the BNS merger GW170817 
has launched the new field of EM-GW multi-messenger astronomy~\cite{Abbott2017_multimessenger}.
Ever since the breakthrough simulations of BBHs in numerical 
relativity about a decade ago~\cite{Pretorius2005,Campanelli2006,Baker2006}, increasingly more
accurate models of BBH merger waveforms across the source
parameter space have been generated~\cite{Mroue2013,Jani2016,Healy2017_catalog}.  
Together with approximate gravitational wave-form models (see, e.g., 
\cite{Hannam:2013oca,Schmidt:2014iyl,Khan:2015jqa, 
2014PhRvD..89f1502T,Pan:2013rra,Babak:2016tgq}), these numerical relativity simulations 
played a crucial role in the parameter estimation of GWs~\cite{Aasi2014}
by LIGO-VIRGO \cite{Heal:2017abq,LovelaceLousto2016,Lange:2017wki}.

Among the missions of current and future GW detectors are tests of General 
Relativity (GR)~\cite{Abbott2016_GW150914_testGR}. 
While the remnant black hole (BH) mass and spin can be estimated from the 
inspiral phase~\cite{Lousto2010}, measuring the quasinormal 
ringdown~\cite{Teukolsky1974,Chandrasekhar1975,Nollert1999,Berti2009} 
of the remnant BH in the GW signal 
will provide an independent measurement of its mass and spin~\cite{Echeverria1989}, 
as well as tests of the no-hair theorem and GR~\cite{Dreyer2004,Gossan2012}. 
Accurate modeling of the ringdown of a highly distorted remnant Kerr BH after merger is 
only possible using numerical relativity simulations of BBH coalescence through merger.
 
Arguably, one of the most widely used evolution schemes for these type of simulations
is the Baumgarte-Shapiro-Shibata-Nakamura (BSSN) 
formulation~\cite{Shibata1995,Baumgarte1999}.\footnote{Sometimes referred to as BSSNOK, 
because it is based on the strategy of~\cite{Nakamura1987} to simplify the spatial Ricci tensor.}
It is based on the Arnowitt-Deser-Misner (ADM) formulation of Einstein's 
equations~\cite{Dirac1949,Arnowitt1962,Arnowitt2008}, and, like the ADM formulation, 
adopts a $3+1$ foliation of spacetime~\cite{Darmois1927}. Unlike the ADM formulation 
it also introduces a conformal-traceless decomposition as well as conformal connection 
functions (see also~\cite{Baumgarte2010} for a textbook introduction).

To date, most numerical codes that adopt the BSSN formulation use finite 
differencing as well as Cartesian coordinates.
While Cartesian coordinates offer distinct advantages (most importantly, they are 
regular everywhere and do not feature any coordinate singularities),  
there are also several shortcomings: BHs, neutron stars, accretion
disks, etc., are often approximately 
spherical or axisymmetric, and Cartesian coordinates are not well suited 
to take advantage of these approximate symmetries. Furthermore, Cartesian
coordinates over-resolve angular directions at large distances,
which leads to the necessity of employing box-in-box mesh refinement. 

In large part, the main target of vacuum numerical relativity simulations are BBH
mergers, whose remnants are
Kerr~\cite{Kerr1963} BHs. Being axisymmetric or nearly axisymmetric,
these merger remnants are prime targets for evolutions in spherical
coordinates. 
In~\cite{Montero2012b}, the authors implemented the BSSN equations
in spherical coordinates without regularization in spherical symmetry.
An important ingredient in obtaining stability was the use of the partially 
implicit Runge-Kutta methods developed in~\cite{CorderoCarrion:2012ic}, even 
though it became clear later that stability can also be achieved with higher-order 
fully explicit Runge-Kutta methods~\cite{Cordero-Carrion:2014}.
In~\cite{Baumgarte2013}, the authors extended the evolution system described 
in~\cite{Montero2012b} to full 3D and performed the first numerical relativity 
simulations in spherical coordinates without the assumption of any symmetries. 
The key idea in this approach is to handle the 
coordinate singularities at the origin and on the axis analytically, rather than 
numerically, which can be achieved with the help of a reference-metric formulation 
of the BSSN equations~\cite{Bonazzola2004,Shibata2004,Brown2009b,Gourgoulhon2012,Montero2012b} 
together with a proper rescaling of all tensorial variables. The same methods can also be 
applied to relativistic hydrodynamics~\cite{Montero2014}. Several examples of vacuum and 
hydrodynamics simulations in spherical coordinates, including an off-center BH head-on 
collision, can be found in~\cite{Baumgarte2015}, and simulations of critical collapse in 
the absence of spherical symmetry in~\cite{Baumgarte:2015aza,Baumgarte2016,Gundlach2016,Gundlach2017}.
This approach has been generalized in the \SENR code~\cite{Ruchlin:2018com,SENRNRPy:web} for 
various other curvilinear coordinate systems.

The use of spherical coordinates has clear advantages. Most importantly, the grid can 
take advantage of the approximate symmetries of the astrophysical objects to be simulated.  
Also, the number of angular grid points is independent of radius, while in Cartesian 
coordinates the number of points per great circle grows with distance from the origin. 
The unigrid (i.e. single computational domain without mesh refinement)
character of a spherical mesh does not produce short-wavelength noise as is the case 
for simulations with mesh refinement boundaries~\cite{Zlochower:2012fk}. 
From a computational standpoint, our unigrid implementation in the \ET offers 
another advantage: It is well documented that mesh refinement codes do not scale as well
as unigrid codes (see~\cite{Schnetter2013} for comparing scaling properties of 
unigrid and mesh refinement in the \ET).

However, these advantages compared to Cartesian coordinates come at a price: Spherical
coordinates have a well know limitation in the form of severely shorter time steps
due to the Courant-Friedrichs-Lewy (CFL) condition, as the cell volumes are not constant,
but decrease with increasing latitude towards the pole and decreasing radius towards the 
origin. A related issue is that the coordinate system becomes singular both at the 
origin and the polar axis, where coordinate values become multivalued. 

An approach to combine the best of both worlds is the use of multipatch
computational domains, in which the domain is broken up into several overlapping
patches locally adapted to the underlying symmetries of the physical system and
free of coordinate singularities and the time step limitations of spherical unigrid 
meshes~\cite{Ronchi1996,Zink2008,Fragile2009a,Reisswig2013,Kageyama2004,
Wongwathanarat2010,Melson2015,Shiokawa2017}.
The SpeC code~\cite{SpEC} uses such a 
multipatch grid structure~\cite{Szilagyi:2009qz}, but in the context
of a pseudospectral evolution scheme. Other  techniques in numerical
relativity codes for dealing with
the polar singularities  are the use of stereographic angular
coordinates, coupled to the eth formalism~\cite{Gomez:1996ge}, as is done in the PITT Null
code~\cite{Bishop:1997ik}, and the use of {\it cubed spheres}, as is
done in the {\sc Llama} infrastructure~\cite{Pollney:2009yz}.
The use of multipatch grids, however, is not free of caveats either: Interpatch 
boundaries require interpolation of fields in ghost zones
which might introduce similar numerical noise as Cartesian mesh-refinement boundaries.

In this work, we report on an implementation of the BSSN equations in spherical coordinates 
described in~\cite{Baumgarte2013} as a thorn called {\sc SphericalBSSN} in the publicly available 
\ET~\cite{ET,Loffler2012}, using code for the BSSN equations provided by~\cite{Ruchlin:2018com}.  
The \ET was designed with Cartesian coordinates in mind, so that we had to adapt 
our implementation of spherical coordinates to its infrastructure in some regards. 
We first identify the $x$, $y$ and $z$ coordinates 
defined in the \ET with $r$, $\theta$ and $\varphi$. The \ET uses a vertex-centered grid
for finite differencing, meaning that grid points are placed on the edges of the 
physical domain.  This is not 
desirable in spherical coordinates, because grid points at the origin or on the axis 
would be singular.  We therefore move both the $r$ and $\theta$ axes by half a grid point, 
so that, effectively, we implement a cell-centered grid in these directions (compare 
Fig.~1 in~\cite{Baumgarte2013}). While, in Cartesian coordinates, the domain boundaries 
in $x$, $y$ and $z$ all correspond to outer boundaries, only the upper $r$ domain 
boundary corresponds to an outer boundary in spherical coordinates. All other boundary 
conditions are \enquote{inner} boundary conditions. For the $\varphi$ coordinate, these 
boundary conditions result from periodicity, while for the $\theta$ direction as well 
as at $r = 0$, the boundary conditions result from parity across the pole or the origin.  
For all inner boundaries, the ghost zones are filled in using properly identified 
interior grid points (see again Fig.~1 in \cite{Baumgarte2013} for an illustration), 
taking into account the parity of tensorial quantities.

In MPI-parallelized domain decompositions, the inner boundary conditions can require 
information from different processes and are therefore more difficult to implement than
in the context of OpenMP-parallelized, single-domain implementations. Using the 
{\sc Slab} thorn~\cite{Loffler2012}, we have
implemented the inner boundary conditions in an MPI-parallelized way, allowing
for arbitrary MPI domain decompositions. 
We also made several changes to existing diagnostics in the \ET so that they can be 
used for evolutions in spherical coordinates, specifically the apparent horizon (AH) 
finder~\cite{Thornburg:2003sf, Schnetter:2004mc} and a thorn that 
computes quasilocal quantities~\cite{Dreyer:2002mx,Schnetter:2006yt} on AHs.
We test the new thorn, together with the changes in the existing diagnostics, 
for a single Bowen-York spinning BH~\cite{Bowen1980}, which is equivalent to a Kerr BH 
with an axisymmetric Brill wave.

Throughout this paper we use units in which $c=G=1$, Latin indices run 
from 1 to 3, and the Einstein summation convention is used. 

\section{Implementation in the Einstein Toolkit}
\label{sec:Implementation}
The \ET is an open source code suite for relativistic astrophysics 
simulations. It uses the modular {\sc Cactus} framework~\cite{Cactus} 
(consisting of general modules called \enquote{thorns}) 
and provides adaptive mesh refinement (AMR) via the {\sc Carpet}
driver~\cite{Carpet,Goodale2003,Schnetter2004}. Here we describe the implementation of the BSSN
evolution code in spherical coordinates within the Toolkit. All codes
mentioned here are either publicly available already, or are in the process of being 
released. 

\subsection{Evolution system}
\label{sec:evolution}
As outlined in~\cite{Montero2012b,Baumgarte2013,Baumgarte2015}, the key 
idea in allowing stable evolutions of
the BSSN~\cite{Nakamura1987,Shibata1995,Baumgarte1999} equations in spherical coordinates 
is to treat the coordinate singularities analytically rather than numerically.  
Specifically, the equations contain terms that diverge with ${\cal O}(r^{-2})$ close to 
the origin and ${\cal O}(\sin^{-2} \theta)$ close to the axis. Adopting a reference-metric 
formulation of the BSSN equations~\cite{Bonazzola2004,Shibata2004,Brown2009b,Gourgoulhon2012,Montero2012b} 
together with a proper rescaling of all tensorial quantities, these terms can be 
differentiated analytically, and, for regular spacetimes, all code variables remain finite.
We also assume absence of conical singularities, which is sometimes 
referred to as ~\enquote{elementary flatness}~\cite{Stephani:2003tm}.
This approach has been generalized in~\cite{Ruchlin:2018com} for a larger number of 
curvilinear coordinate systems, such as spherical coordinates with a $\sinh(r)$ radial
coordinate and cylindrical coordinates, among others. 
We give a summary of the evolution system below, and refer the reader to the full details
in~\cite{Baumgarte2013,Baumgarte2015,Ruchlin:2018com}.

Central to the method is  the conformally related spatial metric
\begin{equation}
\bar{\gamma}_{ij} = e^{-4\phi} \gamma_{ij},
\end{equation}
where $\gamma_{ij}$ is the physical spatial metric, and $\phi$ the conformal factor 
\begin{equation}
e^{4\phi} = (\gamma / \bar{\gamma})^{1/3},
\end{equation}
where $\gamma$ and $\bar{\gamma}$ are the determinants of the physical and conformally 
related metric, respectively. In order to make the conformal rescaling unique, we adopt 
Brown's \enquote{Lagrangian} choice~\cite{Brown2009b}
\begin{equation}
\partial_t \bar{\gamma} = 0,
\end{equation}
fixing $\bar{\gamma}$ to its initial value throughout the evolution.
Similarly, the conformally related extrinsic curvature is defined as
\begin{equation}
\bar{A}_{ij} = e^{-4\phi}\left(K_{ij}-\frac{1}{3}\gamma_{ij}K\right),
\end{equation}
where $K_{ij}$ is the physical extrinsic curvature and $K=\gamma^{ab}K_{ab}$ its trace.

The main idea is to write the conformally related metric as the sum of the flat background
metric plus perturbations (which need not be small)
\begin{equation}
\bar{\gamma}_{ij} = \hat{\gamma}_{ij} + \epsilon_{ij},
\end{equation}
where $\hat{\gamma}_{ij}$ is  the reference metric in spherical coordinates,
\begin{equation}
\hat{\gamma}_{ij} = \begin{pmatrix}
1 & 0 & 0 \\
0 & r^2 & 0 \\
0 & 0 & r^2 {\rm sin}^2 \theta
\end{pmatrix},
\end{equation}
and the corrections $\epsilon_{ij}$ are given by
\begin{equation}
\epsilon_{ij} = \begin{pmatrix}
h_{rr} & r\, h_{r\theta} & r\, {\rm sin} \theta\, h_{r\varphi} \\
r\, h_{r\theta} & r^2\, h_{\theta \theta} & r^2\, {\rm sin} \theta\, h_{\theta \varphi} \\
r\, {\rm sin} \theta\, h_{r\varphi} & r^2\, {\rm sin} \theta\, h_{\theta \varphi} & r^2\, {\rm sin}^2 \theta\, h_{\varphi \varphi}
\end{pmatrix},
\end{equation}
where $h_{ij}$ is the rescaled evolved metric. This idea is similar to 
bimetric formalisms~\cite{Rosen1963,Cornish1964,Rosen1973,NahmadAchar1987,Katz1985,
Katz1988,Katz1990} in GR, in which reference metrics are
employed to give physical meaning to pseudotensors in curvilinear coordinates,
or in the integration of the Ricci scalar on a hypersurface~\cite{Gourgoulhon1994}.
The evolved conformally rescaled extrinsic 
curvature $a_{ij}$ is similarly related to the conformally related extrinsic curvature 
$\bar{A}_{ij}$
\begin{equation} \label{aij}
\bar{A}_{ij} = \begin{pmatrix}
a_{rr} & r\, a_{r\theta} & r\, {\rm sin} \theta\, a_{r\varphi} \\
r\, a_{r\theta} & r^2\, a_{\theta \theta} & r^2\, {\rm sin} \theta\, a_{\theta \varphi} \\
r\, {\rm sin} \theta\, a_{r\varphi} & r^2\, {\rm sin} \theta\, a_{\theta \varphi} & r^2\, {\rm sin}^2 \theta\, a_{\varphi \varphi}
\end{pmatrix}.
\end{equation} 
The conformal connection coefficients $\bar{\Lambda}^i$ are treated as independent variables that
satisfy the initial constraint
\begin{equation}
\bar{\Lambda}^i - \Delta^i = 0.
\end{equation} 
Here 
\begin{equation}
\Delta^i \equiv \bar{\gamma}^{ab}\Delta^i_{ab}
\end{equation}
and $\Delta^i_{jk}$ is the difference between the Christoffel symbols of the
conformally rescaled and flat reference metric,
\begin{equation}
\Delta^i_{jk} \equiv \bar{\Gamma}^i_{jk} - \hat{\Gamma}^i_{jk}.
\end{equation}
The conformal connection coefficients $\bar{\Lambda}^i$ therefore transform like vectors 
in the reference-metric formalism. Similar to our treatment of the metric and the extrinsic 
curvature we write 
\begin{equation} \label{Lambda}
\bar{\Lambda}^i = \begin{pmatrix}
\lambda^r \\
\lambda^\theta / r \\
\lambda^\varphi / (r \sin \theta) 
\end{pmatrix}
\end{equation}
and evolve the variables $\lambda^i$ in our code. We refer the reader 
to~\cite{Baumgarte2013,Baumgarte2015,Ruchlin:2018com} for the full details of the evolution 
system.

The physical metric $\gamma_{ij}$ and the physical extrinsic curvature 
$K_{ij}$ can be reconstructed from the evolved variables $h_{ij}$ and $a_{ij}$ as follows:
\begin{equation}
\gamma_{ij} = e^{4\phi}\begin{pmatrix}
1+h_{rr} & r\, h_{r\theta} & r\, {\rm sin} \theta\, h_{r\varphi} \\
r\, h_{r\theta} & r^2\, (1+h_{\theta \theta}) & r^2\, {\rm sin} \theta\, h_{\theta \varphi} \\
r\, {\rm sin} \theta\, h_{r\varphi} & r^2\, {\rm sin} \theta\, h_{\theta \varphi} & r^2\, {\rm sin}^2 \theta\, (1+h_{\varphi \varphi})
\end{pmatrix},
\end{equation}
and
\begin{equation}
K_{ij} = e^{4\phi} \bar{A}_{ij} + \frac{1}{3} \gamma_{ij} K.
\end{equation}
Together with the lapse $\alpha$ and the shift $\beta^i$, this set of the $3+1$ 
variables $\{\alpha,\beta^i,\gamma_{ij},K_{ij}\}$, expressed in spherical coordinates,
is stored in the thorn {\sc ADMBase} to interface with existing diagnostics in 
the \ET~\cite{ET,Loffler2012}. 

Numerical code for the evolution system is provided by the \SENR code, and the time
integration is performed with the method of lines as implemented in the 
{\sc MoL}~\cite{ET,Loffler2012} thorn.

\subsection{Spherical parity boundary conditions}
\label{sec:BC}
There is no global and regular one-to-one map from spherical to Cartesian coordinates.  
Instead, at least two charts are needed to cover an entire sphere of a given radius. 
Ultimately, this is due to the fact that $\theta$ and $\varphi$ are multivalued at the 
coordinate origin, and $\varphi$ is multivalued at the polar axis. As a result, 
the Jacobian from spherical to Cartesian coordinates diverges both at the origin and 
polar axis. 
In our implementation in the \ET we use the existing Cartesian grid infrastructure as 
spherical coordinates by implementing the internal boundary conditions in a way that uses
the underlying topologically Cartesian grid.

As explained already in the Introduction, we start by identifying the internal 
$(x,y,z)$ coordinate representation used in {\sc Carpet} with the spherical coordinates 
$(r,\theta,\varphi)$.  {\sc Carpet} uses a vertex-centered grid structure, meaning that 
grid points exist on the edges of the physical domains.  This is not desirable in 
spherical coordinates, because of the coordinate singularities at the origin, $r = 0$, 
and the poles at $\theta = 0$ and $\theta = \pi$.  We therefore shift both the $r$ and 
$\theta$ axes by half a grid point. Therefore, the physical 3D domain has the following 
extents: 
\begin{eqnarray}
r &\in& \left[\frac{dr}{2},r_{\rm max}\right], \\
\theta &\in& \left[\frac{d \theta}{2},\pi-\frac{d \theta}{2}\right], \\
\varphi &\in& \,\left[0,2\pi - d \varphi \right].
\end{eqnarray}  
Effectively, we therefore adopt a cell-centered grid in the $r$ and $\theta$ directions, 
but maintain a vertex-centered grid in the $\varphi$ direction.

Cartesian coordinates are topologically $\mathbf{R}^3$, and all domain boundaries 
for large or small values of the coordinates $x$, $y$ or $z$ correspond to outer boundaries.  
In spherical coordinates, on the other hand, only $r_{\rm max}$ corresponds to an 
outer boundary, while all other domain boundaries represent \enquote{inner boundaries}.   
At $r = 0$, for example, a radial grid line can be extended to negative values of $r$.  
We allow for $n_{\rm gzr}$ ghost zone grid points at negative $r$; these ghost zone 
grid points correspond to interior grid points with positive $r$ for some other values 
of the angles $\theta$ and $\varphi$ (see Fig.~1 in~\cite{Baumgarte2013} for an 
illustration). Specifically, we identify ghost zones at the origin with interior 
grid points at the coordinate locations
\begin{eqnarray}
r & \to & - r \\
\theta & \to & \pi - \theta, \\
\varphi & \to & \varphi + \pi.
\end{eqnarray}
We can then fill these ghost zones by applying internal parity boundary conditions, 
which we explain in more detail below. Similarly, meridians, i.e., great circles 
of constant $\varphi$, can be extended across the pole, and the ghost zones there 
can again be identified with internal grid points. For $\theta_{\rm min} = 0$ we have  
\begin{eqnarray}
\theta & \to & - \theta, \\
\varphi & \to & \varphi + \pi,
\end{eqnarray}
and for $\theta_{\rm max} = \pi$ 
\begin{eqnarray}
\theta & \to & \pi - \theta, \\
\varphi & \to & \varphi + \pi.
\end{eqnarray}
We also introduce ghost zones for 
$\varphi$, which can be filled by imposing periodicity. We note that application 
of this scheme requires an even number of grid points in the $\varphi$ direction.

For scalar quantities, field values at interior grid points can be copied directly into 
the corresponding ghost zone grid points. For tensorial quantities, however, we 
have to take into account the fact that the direction of unit vectors changes when 
crossing the origin or pole (see~\cite{Baumgarte2013,Ruchlin:2018com} for more details).  
This observation leads to parity factors that arise in the application of the inner 
boundary conditions. We list these factors in Table~\ref{table:parities}.

By using the cell-centered grid in $r$ and $\theta$, and using the described 
internal boundary conditions, we are able to have a one-to-one mapping of the 
internal Cartesian coordinates in the \ET, which are topologically $\mathbf{R}^3$,
with the spherical coordinates used in the evolution.

\begin{table}
\begin{tabularx}{\columnwidth}{ZZZ}
&Origin &Axis \\
\hline
$V_r$ &-- &+ \\
$V_{\theta}$ &+ &-- \\
$V_{\varphi}$ &-- &-- \\
\hline
$T_{rr}$ &+ &+ \\
$T_{r\theta}$ &-- &-- \\
$T_{r\varphi}$ &+ &-- \\
$T_{\theta\theta}$ &+ &+ \\
$T_{\theta\varphi}$ &-- &+ \\
$T_{\varphi\varphi}$ &+ &+ \\
\end{tabularx}
\caption{Table showing the spherical parity factors for vectors and tensors 
(see Table I in~\cite{Baumgarte2013}).}
\label{table:parities}
\end{table}

\begin{figure}[thb]

\centering
    \resizebox{\columnwidth}{!}{
    \includegraphics[scale=1.0]{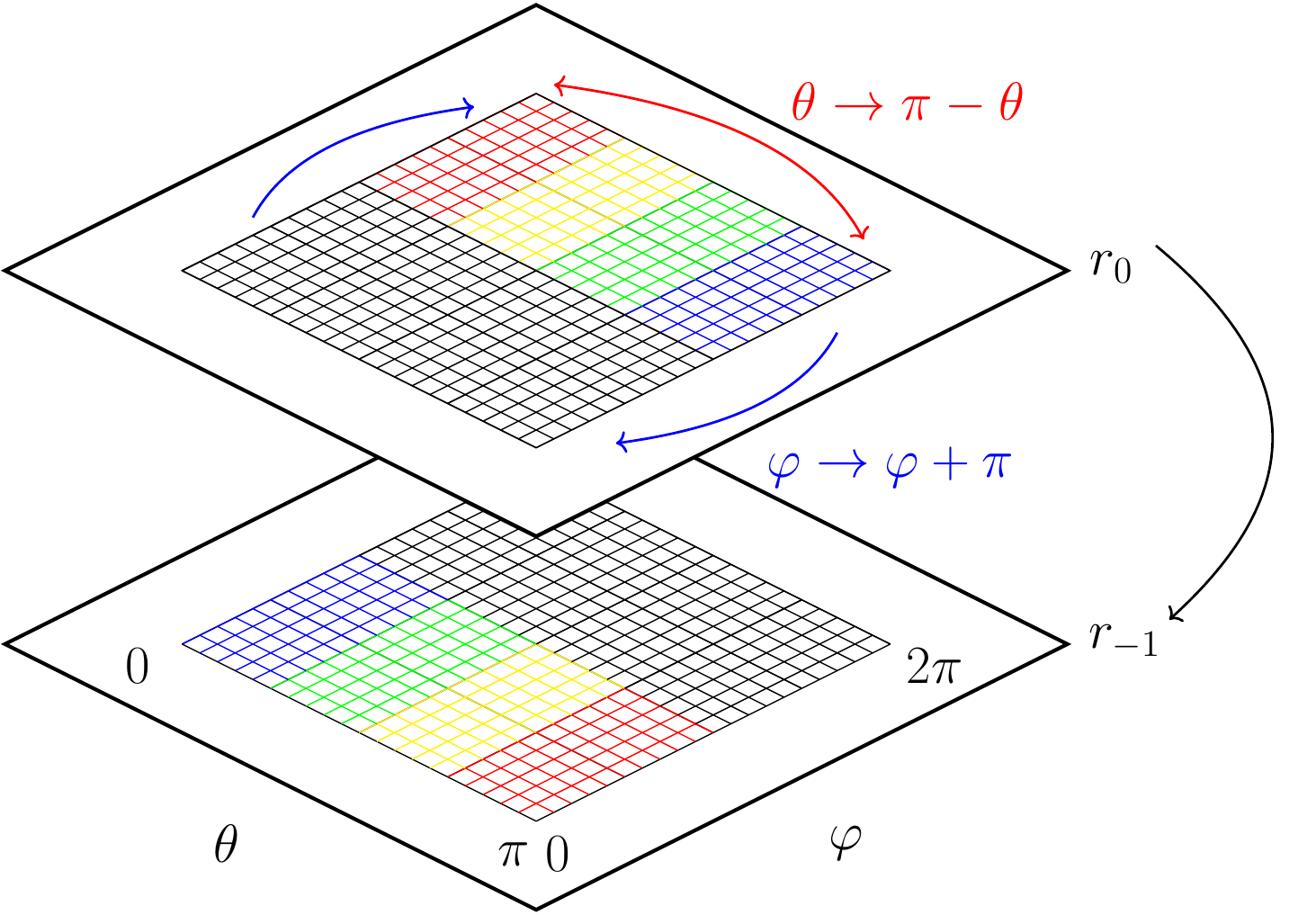}
}
\caption{Diagram depicting the slab transfers involved in the $r_{\rm min}$
boundary condition.}
\label{fig:rBC}
\end{figure}

\begin{figure}[thb]

\centering
    \resizebox{\columnwidth}{!}{
    \includegraphics[scale=1.0]{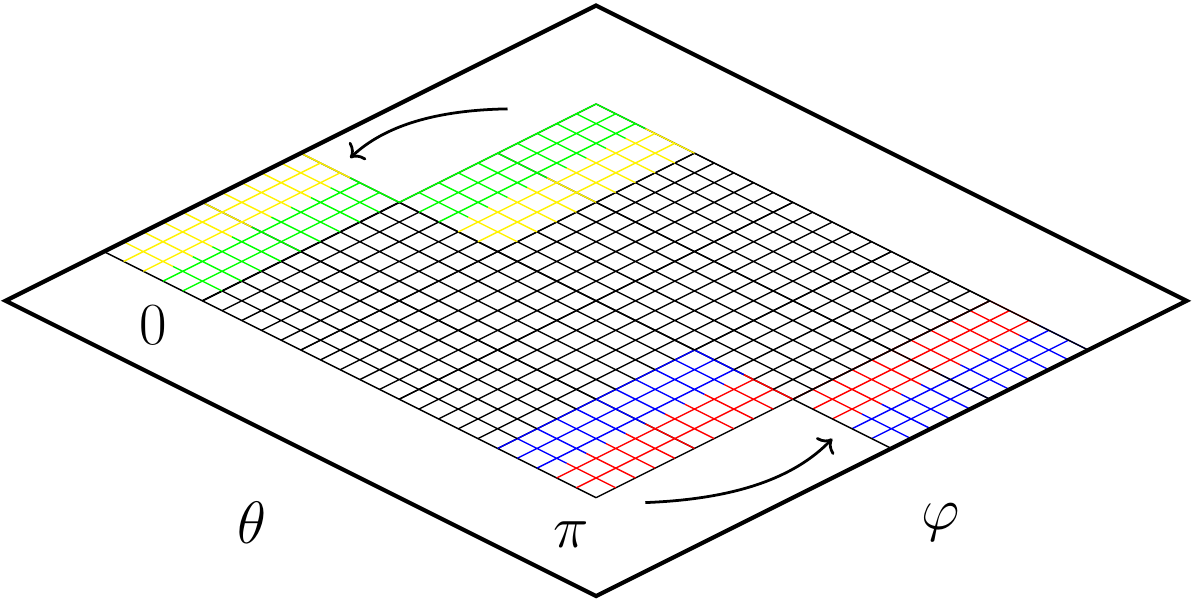}
}
\caption{Diagram showing the slab transfers involved in the $\theta_{\min}$ 
and $\theta_{\rm max}$ boundary conditions.}
\label{fig:thetaBC}
\end{figure}

Allowing for arbitrary MPI domain decompositions requires communication across
processes, as a given process might not be in possession of the point that is mapped to 
a ghost zone in its domain.  We have implemented these boundary conditions using the 
{\sc Slab} thorn, which provides MPI-parallelized infrastructure to take 3D subarrays 
(\enquote{slabs}) of the 3D domain, manipulate them and then broadcast the manipulated 
slabs back to all processors that contain a part of it.  In what follows, we show how the 
internal boundary conditions are implemented as slab transfers using the
{\sc SLAB} thorn (see Figs.~\ref{fig:rBC} and~\ref{fig:thetaBC}).

The source slab for the boundary condition at the origin contains the first $n_{\rm gzr}$ physical 
points in $r$, where, again, $n_{\rm gzr}$ is the number of ghost zones in the $r$ direction, 
and all physical points 
in $\theta$ and $\varphi$. The operation $\theta  \to  \pi - \theta$
is performed by inverting the slab in the $\theta$ direction, while 
the $\varphi  \to  \varphi + \pi$ operation corresponds to moving all points from 
$[0,\pi] \to [\pi,2 \pi]$ and, taking into account
the periodicity in $\varphi$, all points from $[\pi,2 \pi] \to [0,\pi]$. The $\varphi$ part of 
the operation is achieved using two separate calls to the slab transfer. Finally, the
source slab is inverted in $r$ and the resulting slabs transferred into the ghost zones of
the domain. This is illustrated in Fig.~\ref{fig:rBC}. 
Note that the procedure only fills ghost zones that correspond to physical points in 
$\theta$ and $\varphi$, while \enquote{double} and \enquote{triple} ghost zones 
on the edges and corners of the computational grid need to be filled by subsequent
internal boundary conditions.

We then proceed by imposing the $\theta$ boundary conditions in a similar manner (see
schematic in Fig.~\ref{fig:thetaBC}), followed by 
applying periodic boundary conditions in $\varphi$ using an existing thorn in the 
\ET. This order ensures that all ghost zones 
that need to be specified by internal parity boundaries are filled correctly. 

In future applications that include magnetohydrodynamics and/or other matter fields, 
the same parity boundary conditions will apply to the matter fields as well.  
We have therefore implemented these boundary conditions in a separate thorn, 
{\sc SphericalBC}, so that they are available for all evolved quantities, and not 
only for the spacetime evolution.

\subsection{Time Step considerations}
\label{sec:time_step}
It is well known that the time integration of hyperbolic PDEs in spherical coordinates 
suffers from severe CFL time step restrictions, as the cell sizes
become smaller and smaller with increasing latitude from the equator
towards the poles, and decreasing distance from the origin. In a spherical unigrid in flat spacetimes, 
the time step due to the CFL condition is given 
by~\cite{Baumgarte2013}:
\begin{equation}
dt = \mathcal{C}\, {\rm min}\left[d r,\frac{d r}{2} d \theta,
\frac{d r}{2} \sin(\frac{d \theta}{2})d \varphi \right],
\end{equation} 
where the CFL factor $\mathcal{C}$ is chosen between 0 and 1. The time step is therefore 
limited by the azimuthal distance between cells at the origin and polar axis.
Compared with Cartesian coordinates, where $dt \approx dx$ (when the same 
spatial resolution $dx$ is used in all three coordinates), the time step in spherical
coordinates varies as $dt \approx d r \, d \theta \, d \varphi$.  
Thus high angular resolution will impose severe time step restrictions in spherical
coordinates. 

When using a high number of azimuthal cells, the CFL 
restriction might render the numerical integration 
computationally unfeasible. There are several approaches to mitigate this problem
(for an introduction, see e.g.~\cite{Boyd2001}), from various {\it multipatch} 
approaches~\cite{Ronchi1996,Gomez:1996ge,Bishop:1997ik,Zink2008,Fragile2009a,Reisswig2013,Kageyama2004,
Wongwathanarat2010,Melson2015,Shiokawa2017} to reducing the number of azimuthal cells at high 
latitudes as mesh coarsening in the azimuthal direction~\cite{Liska2018}, 
focusing resolution of the polar angle at the equator~\cite{Korobkin2011,Noble2012},  
or the use of filters~\cite{Shapiro1970,Gent1989,Jablonowski2004,Mueller2015}, to name a few. 

To circumvent the severe time step limitation in cases when evolving BHs centered 
at the coordinate origin, we have devised a simple excision strategy in order to enlarge
time steps in these evolutions.
Specifically, we employ a radial extrapolation of all evolved variables deep inside 
the AH during the evolution, which essentially amounts to excising parts of the BH interior. 
Similar strategies have been employed and shown to work in the context of puncture
BH~\cite{Etienne2007,Brown2007,Brown2009}.
Within a fixed number of radial points the BSSN variables are not evolved but rather
linearly extrapolated inwards radially from the first evolved points. When using this
technique, we get a dramatic increase in time step, which is now given by:
\begin{equation}
dt = \mathcal{C}\, {\rm min} \left[dr, A dr d\theta,
A dr \sin(\frac{d \theta}{2}) d \varphi \right],
\end{equation}
where $A \equiv n_{\rm Ex}-1+\frac{1}{2}$ and $n_{\rm Ex}$ is the number of
\enquote{excised} radial grid points. In the simulations presented in Sec.~\ref{sec:Results}
below, we choose this parameter such that $n_{\rm Ex} dr \approx r_{\rm AH}/2$ 
initially, and a CFL factor $\mathcal{C} = 0.4$.  
We emphasize that we employ this excision for the purposes of speeding up the 
simulation only -- it is not needed for stability. In Fig.~\ref{fig:ex_comparison} below 
we compare simulations with and without this excision; we also note that the simulations
in~\cite{Baumgarte2013,Baumgarte2015,Baumgarte:2015aza,Baumgarte2016,Gundlach2016,Gundlach2017,Ruchlin:2018com}
did not use such an excision.

\subsection{Diagnostics}
\label{sec:diagnostics}
We use the {\sc AHFinderDirect} thorn~\cite{Thornburg:2003sf, Schnetter:2004mc} to find
AHs~\cite{Thornburg2007a} and the
{\sc QuasiLocalMeasures} thorn~\cite{Dreyer:2002mx,Schnetter:2006yt} to calculate the angular momentum
of the apparent horizon during the evolution. The BH spin is measured using the 
flat space rotational Killing vector method~\cite{Campanelli2007} that was shown
to be equivalent to the Komar angular momentum~\cite{Komar1959} in foliations adapted
to the axisymmetry of the spacetime~\cite{Mewes2015}.
Both thorns were written explicitly for the Cartesian coordinates employed in 
{\sc Carpet} --- interpolating entirely in the Cartesian basis both the 
{\sc ADMBase} variables $\{\gamma_{ij}, K_{ij}, \alpha, \beta^i\}$
and the partial derivatives of the spatial metric and extrinsic
curvature. As indicated in Sec.~\ref{sec:evolution} above, 
the {\sc ADMBase} variables in the {\sc SphericalBSSN}
thorn are the physical metric, extrinsic curvature, lapse and shift in spherical 
coordinates, which means we need to transform the {\sc ADMBase} variables and their 
partial derivatives to Cartesian coordinates after interpolation. 
This is required because {\sc AHFinderDirect} expects the computational domain to have
Cartesian topology (i.e., any surface with constant $r$, will not appear to be
closed to {\sc AHFinderDirect}). In its original form, {\sc AHFinderDirect} 
interacts with the rest of the Toolkit by requesting
the interpolation of the metric functions, and their derivatives, at
various points in Cartesian coordinates. To make this work with {\sc
SphericalBSSN}, we modify this behavior by transforming the Cartesian
coordinates to spherical (the necessary Jacobians are provided by aliased functions
defined in {\sc SphericalBSSN}), and then after the interpolation step,
transforming the metric functions from spherical to Cartesian
coordinates using the Jacobian 
$J^{a}{}_{i} \equiv  \frac{\partial x^a}{\partial x^i}$ 
from spherical to Cartesian coordinates according to
\begin{eqnarray}
 \gamma_{ij} &=& J^{a}{}_{i}\, J^{b}{}_{j}\, \gamma_{ab},\nonumber \\
 \gamma_{ij,k} &=& J^{a}{}_{i,c}\, J^{b}{}_{j}\, J^{c}{}_{k}\, \gamma_{ab},\nonumber \\
 &+& J^{a}{}_{i}\, J^{b}{}_{j,c}\, J^{c}{}_{k}\, \gamma_{ab},\nonumber \\
 &+& J^{a}{}_{i}\, J^{b}{}_{j}\, J^{c}{}_{k}\, \gamma_{ab,c},\nonumber \\
 K_{ij} &=& J^{a}{}_{i}\, J^{b}{}_{j}\, K_{ab},\nonumber \\
 K_{ij,k} &=& J^{a}{}_{i,c}\, J^{b}{}_{j}\, J^{c}{}_{k}\, K_{ab},\nonumber \\
 &+& J^{a}{}_{i}\, J^{b}{}_{j,c}\, J^{c}{}_{k}\, K_{ab},\nonumber \\
 &+& J^{a}{}_{i}\, J^{b}{}_{j}\, J^{c}{}_{k}\, K_{ab,c},\nonumber \\
 \beta^i &=& J^{i}{}_{a}\, \beta^a,
\end{eqnarray}
where we adopt the convention that indices $a$ and $b$ refer to spherical coordinates 
$r$, $\theta$ and $\varphi$, and indices $i$ and $j$ to the Cartesian 
coordinates $x$, $y$ and $z$ in coordinate transformations, and a comma 
indicates ordinary partial differentiation.
When the origin of the {\sc AHFinderDirect} internal six-patch system coincides with the
coordinate origin, we add a small offset in $\theta$ at points located on the $z$-axis,
as the Jacobian diverges at those points.  

We extract GWs by computing the Weyl scalar $\Psi_4$ using the electric and magnetic 
parts of the Weyl tensor and constructing the numerical tetrad as described 
in~\cite{Baker2002} in spherical coordinates. The calculation of $\Psi_4$ and the
remaining Weyl scalars is contained in a new thorn called {\sc SphericalWeylScal}.
The multipole expansion of the real and imaginary parts of $\Psi_4$ in spin-weighted 
spherical harmonics~\cite{Thorne1980} is performed on the spherical grid used in the 
evolution. {\sc SphericalWeylScal} performs the multipole expansion after the 
calculation of the Weyl scalars.
 
While the Jacobian for spherical coordinates is simple to implement
directly into the analysis thorns, we coded our
modification to {\sc AHFinderDirect}  and {\sc QuasiLocalMeasures}  so
that they call aliased functions. In this way, both codes can now work
with arbitrary coordinate systems, as the calculation of the Jacobians, etc.,
are handled by auxiliary routines.

\subsection{Initial Data}
\label{sec:ID}
As a demonstration of our methods we show in Sec.~\ref{sec:Results} below an evolution 
of a spinning Bowen-York BH~\cite{Bowen1980}, which describes a perturbed Kerr BH~\cite{Kerr1963}.

Bowen-York data are conformally flat, so that $h_{ij} = 0$ identically, as well as 
$\bar\Lambda^i = 0$.  The data are also maximally sliced, so that $K = 0$.  
The momentum constraint can then be solved analytically for the conformally rescaled 
extrinsic curvature; for rotating Bowen-York BHs, the only non-vanishing component is 
the $r\varphi$ component.  Given the analytical solution for this extrinsic curvature, 
the Hamiltonian constraint can then be solved numerically for the conformal factor $e^\phi$.  
The only non-vanishing component of the extrinsic curvature variables $a_{ij}$ 
defined in (\ref{aij}) is then
\begin{equation}
a_{r\varphi} = \frac{3 e^{-6 \phi} J \sin\theta}{r^3},
\end{equation}
where $J$ is the magnitude of the BH's angular momentum 
(see also exercise 3.11 in \cite{Baumgarte2010}).

In order to allow for future applications with more general sets of initial data 
that may have been prepared in Cartesian coordinates, we do not implement the above 
results directly, but instead use the {\sc TwoPunctures} thorn~\cite{Ansorg2004} to 
set up the data. 
This thorn uses spectral methods to solve the Einstein 
constraints, and interpolates the Cartesian 
{\sc ADMBase} variables onto the computational mesh used 
in the simulation.  We have adapted the thorn to interpolate the Cartesian {\sc ADMBase} 
variables onto the spherical grid points instead. Upon the completion of the interpolation, the metric
$\gamma_{ij}$ and extrinsic curvature $K_{ij}$ are transformed from Cartesian 
to spherical coordinates as described in Section~\ref{sec:diagnostics} above.  
The evolved metric variables $h_{ij}$ are then computed from
\begin{equation}
h_{ij} = e^{-4\phi}{\gamma}_{ij}\, \odot \, \begin{pmatrix}
1& 1/r & 1/(r\, {\rm sin} \theta) \\
1/r & 1/r^2 & 1/(r^2\, {\rm sin} \theta) \\
1/(r\, {\rm sin} \theta) & 1/(r^2\, {\rm sin} \theta) & 1/(r^2\, {\rm sin}^2 \theta)
\end{pmatrix} - \mathbbm{1},
\end{equation}
where $\odot$ indicates the Hadamard product (element-wise matrix multiplication) and
$\mathbbm{1}$ the identity matrix, while the evolved extrinsic curvature variables $a_{ij}$ are 
\begin{equation}
\begin{split}
a_{ij} = &e^{-4\phi}(K_{ij}-\frac{1}{3}{\gamma}_{ij}K) \\
&\odot \, \begin{pmatrix}
1& 1/r & 1/(r\, {\rm sin} \theta) \\
1/r & 1/r^2 & 1/(r^2\, {\rm sin} \theta) \\
1/(r\, {\rm sin} \theta) & 1/(r^2\, {\rm sin} \theta) & 1/(r^2\, {\rm sin}^2 \theta)
\end{pmatrix}.
\end{split}
\end{equation}
We confirmed that these variables agree with the values listed above to within 
truncation error.

We complete the specification of the initial data with choices for the initial lapse 
and shift. We choose an initial shift $\beta^i=0$ and an initial lapse $\alpha = e^{-2\phi}$ 
for the {\sc SphericalBSSN} runs and $\beta^i=0$, $\alpha = 2 r/(M + 2r)$ for 
the comparison \ET runs.

\section{Results}
\label{sec:Results}
\begin{figure}
	\centering
	\includegraphics[scale=1.0]{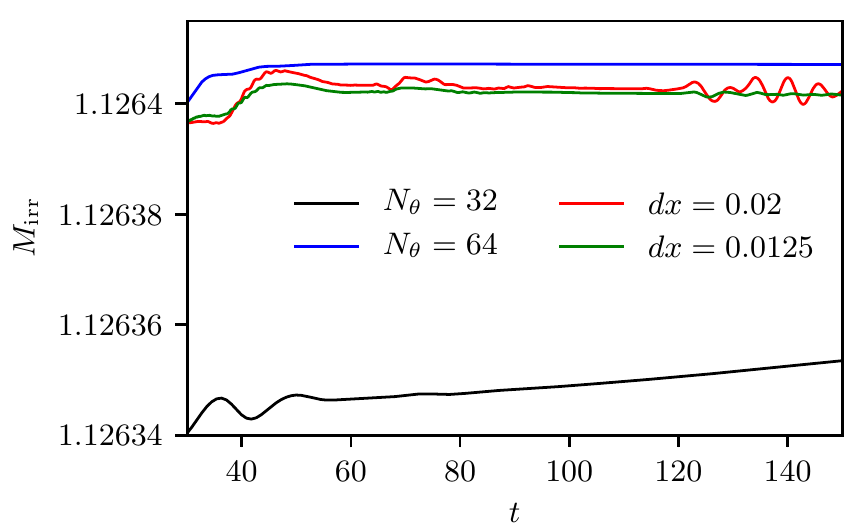}
	\caption{Comparison of the evolution of the irreducible mass
	of the BH for {\sc SphericalBSSN} and {\sc McLachlan} at two different 
	resolutions each.}
	\label{fig:mirr}
\end{figure}

\begin{figure}
	\centering
	\includegraphics[scale=1.0]{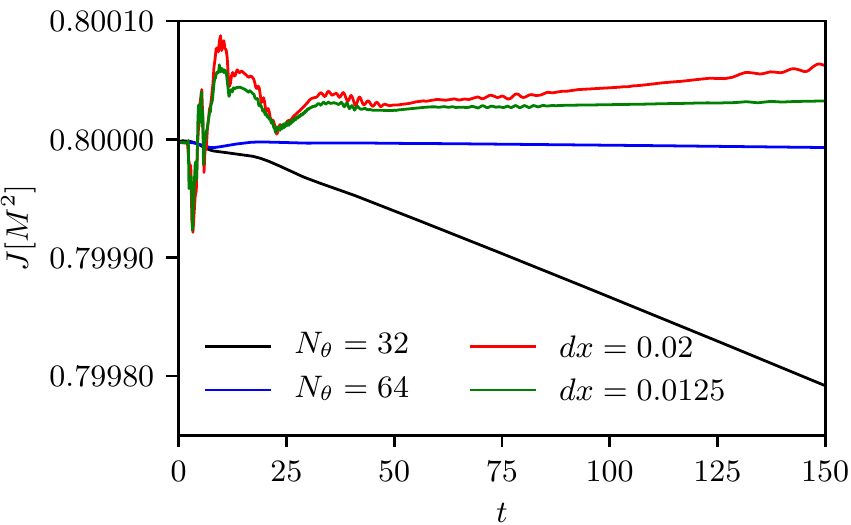}
	\caption{Comparison of the evolution of the AH angular momentum
	for {\sc SphericalBSSN} and {\sc McLachlan} at two different 
	resolutions each.}
	\label{fig:spin}
\end{figure}

\begin{table*}
	\begin{tabular}{c | cc}

	 & {\sc SphericalBSSN} &{\sc McLachlan} \\
	 \hline \\
	 Resolution & $dr=0.02$, $d\theta=\pi/32~(\pi/64)$, $d\varphi=2\pi/4$ & 
	 $dx = dy = dz = 0.02~(0.0125)$ \\
	 Mesh refinement & unigrid & 10 refinement levels~\cite{Carpet} \\
	 Outer boundary & 200 (500) & 512 \\
	 Outer boundary condition & Sommerfeld BC~\cite{Alcubierre2000} & Sommerfeld BC \\
	 FD order & 4th order centered finite differencing & 4th order centered finite differencing \\
	 Upwinding & 4th order upwinding on shift advection terms & 4th order upwinding on shift advection terms\\
	 Kreiss-Oliger dissipation & 5th order dissipation & 5th order dissipation \\
	 Dissipation strength & $\epsilon = 0.1$ & $\epsilon = 0.1$ \\
	 Time integration & Method of lines with RK4 & Method of lines with RK4\\
	 CFL factor $\mathcal{C}$ & 0.4 & 0.4 \\
	 Prolongation & none & 5th order spatial, 2nd order temporal prolongation \\
	 Lapse evolution & 1+log slicing~\cite{Bona1995} & 1+log slicing \\
	 Lapse advection & yes & yes \\
	 Shift evolution &$\Gamma$-driver~\cite{Alcubierre2003}, $\eta=1.0$ & $\Gamma$-driver, $\eta = 2.0$ \\
	 Shift advection & yes & yes \\
	 Evolved conformal factor & $W=e^{-2 \phi}$ ~\cite{Marronetti:2007wz} & $W=e^{-2 \phi}$\\
	 Gravitational wave extraction & $\Psi_4$ with {\sc SphericalWeylScal} & 
	 $\Psi_4$ with {\sc WeylScal4} thorn~\cite{Hinder2011} \\
	 GW extraction radii & {20, 60, 100, 140, 180} & {20, 60, 100, 140, 180} \\

	\end{tabular}
	\caption{Summarizing the main parameters of the simulations performed with {\sc SphericalBSSN}
	and the {\sc McLachlan} thorn.}
	\label{table:sim_details}
\end{table*}
We perform simulations of a single, spinning, and initially conformally flat BH
(the Bowen-York solution~\cite{Bowen1980}), with the initial data prepared as described 
in Section \ref{sec:ID} above. Since the Kerr~\cite{Kerr1963} spacetime is not
conformally flat, these initial data
represent a spinning BH with gravitational wave content that will be radiated 
away~\cite{Gleiser1998}, allowing the BH to settle to the Kerr solution.
In all results presented here, dimensionful quantities are reported in terms of
$M=1$.  The BH has an initial spin $J=0.8M^2$ and an ADM mass~\cite{Arnowitt1962,Arnowitt2008} 
of $M_{\rm BH}=1.18112M$, giving a Kerr parameter $a \equiv J/M_{\rm BH} = 0.677 M$. 
We perform simulations of these initial data using our {\sc SphericalBSSN} implementation.  
For comparison, we evolve the same initial data in Cartesian coordinates
using the {\sc McLachlan}~\cite{Brown2009,Reisswig2011} thorn. {\sc McLachlan} is a finite 
difference code generated using {\sc Kranc}~\cite{Husa2006} 
that solves the BSSN equations as part of the \ET. For our comparisons here we use 
4th-order spatial finite differences in both codes, but we note that \SENR and
{\sc McLachlan} are capable of providing finite-difference stencils for the BSSN equations 
at arbitrary order. We summarize the details of the relevant simulation parameters in 
Table~\ref{table:sim_details}.

\subsection{BH mass and spin}
\label{sec:bh_mass}
In Figs.~\ref{fig:mirr} and~\ref{fig:spin} we plot the evolution of the irreducible mass
of the BH and its angular momentum, respectively. During the first $35 M$ of the evolution 
the BH mass increases due to the absorption of some of the GW content in the 
spacetime (see Fig.~\ref{fig:ex_comparison} below). We omit this initial time in 
Fig.~\ref{fig:mirr}, and instead show the long-term behavior after the BH has settled down.  
We show results for two different $\theta$ resolutions ($N_{\theta} = 32$ and $64$) with 
{\sc SphericalBSSN} and two resolutions ($dx=0.0125$ and $0.02$ on the finest mesh) using 
the \ET in Cartesian coordinates with box-in-box mesh refinement. The radial resolution in 
both evolutions ($dr=0.02$ and $dx=0.02$ on the finest Cartesian mesh which covers the AH) 
gives approximately 25 radial points across the minimum diameter (0.25) of the AH initially. 
For the irreducible mass shown in Fig.~\ref{fig:mirr}, the results obtained with the 
higher resolution {\sc SphericalBSSN} and the two Cartesian runs agree well, while 
the lower resolution {\sc SphericalBSSN} run appears to be under-resolved, showing a 
linear growth in the irreducible mass that is unphysical. 
There is a notable absence of oscillations in the higher resolution {\sc SphericalBSSN} 
run, which can be seen in both Cartesian runs (converging away with increasing 
resolution).
 
The evolution of the angular momentum of the AH, shown in Fig.~\ref{fig:spin}, exhibits a 
similar behavior. The two high-resolution runs with {\sc McLachlan} and {\sc SphericalBSSN}
perform similarly, while the lower resolution runs show linear drifts in both Cartesian 
and spherical coordinates. The Cartesian simulations show larger initial oscillations
that do not seem to converge away with increasing resolution, likely due to reflections 
of short-wavelength modes across mesh boundaries~\cite{Zlochower:2012fk,Etienne:2014tia}. 
Just as for the irreducible mass, the high-resolution {\sc SphericalBSSN} simulations 
performs best. 

To test the effect of the excision described in Section~\ref{sec:time_step}, we plot
the initial evolution of the irreducible mass for a run with, and one
without, our excision procedure in Fig.~\ref{fig:ex_comparison}.   
Evidently, the excision procedure does not have any visible effect on 
the accuracy or stability of our method.
\begin{figure}
	\centering
	\includegraphics[scale=1.0]{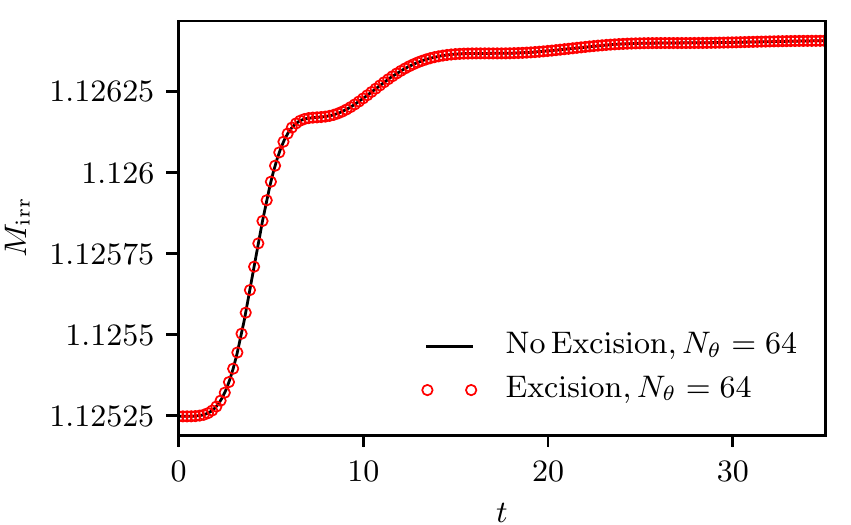}
	\caption{Comparing the evolution of the BH irreducible mass
          for two {\sc SphericalBSSN} runs, with and without the excision
          algorithm described in Sec.~\ref{sec:time_step}.}
	\label{fig:ex_comparison}
\end{figure}

\subsection{Gravitational waves}
\label{sec:GW}
\begin{figure*}
	\centering
	\begin{tabular}{p{.00\textwidth} >{\centering\arraybackslash}
         m{.5\textwidth} >{\centering\arraybackslash}
         m{.5\textwidth} }
	&\IncG[scale=1.0]{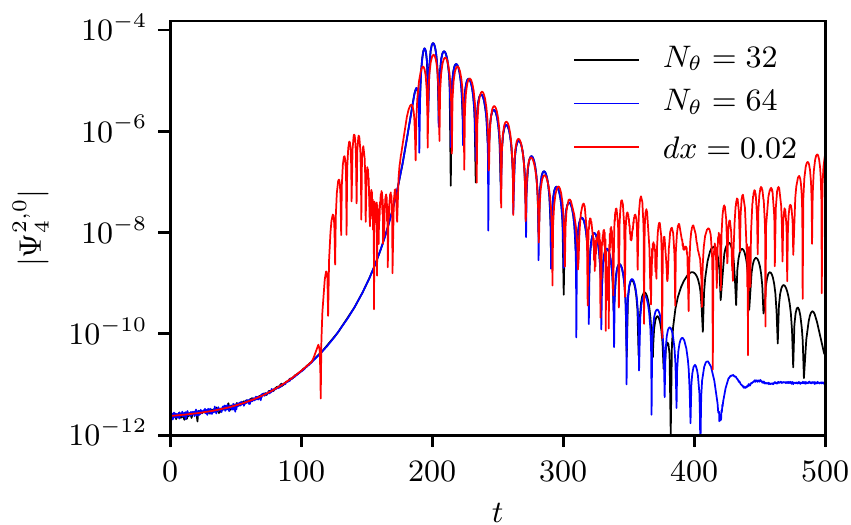}
	&\IncG[scale=1.0]{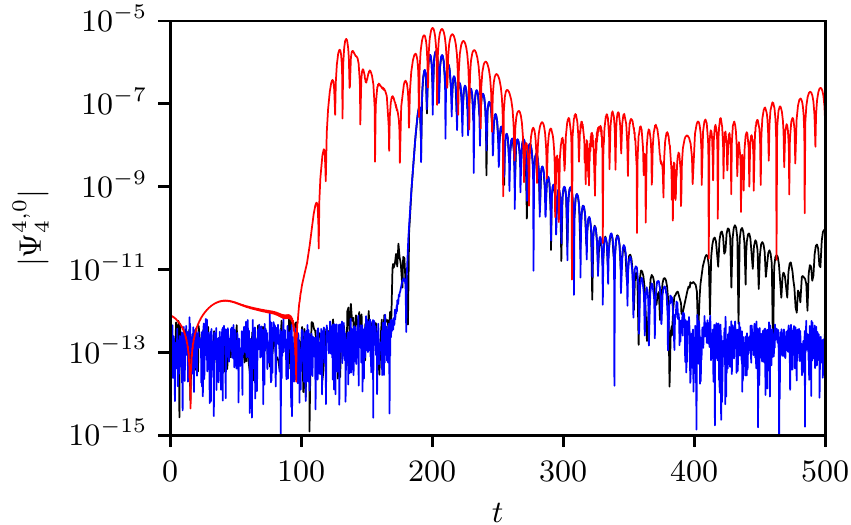}
	\\
	&\IncG[scale=1.0]{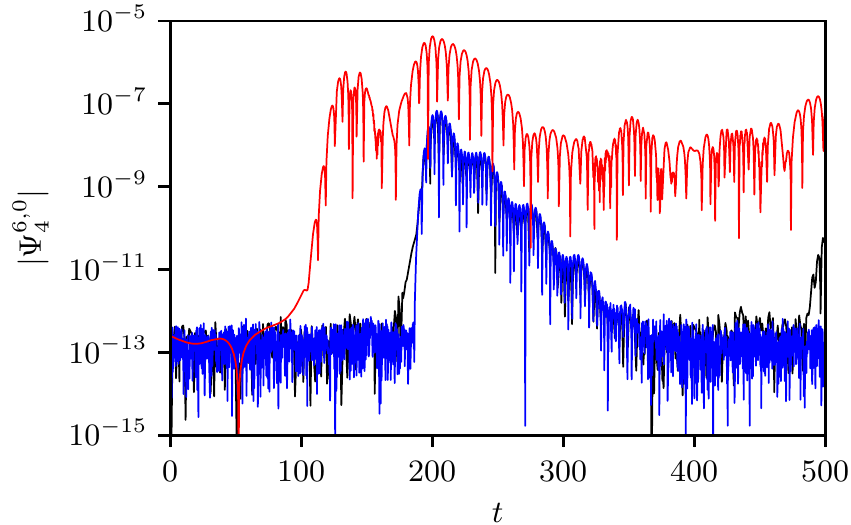}
	&\IncG[scale=1.0]{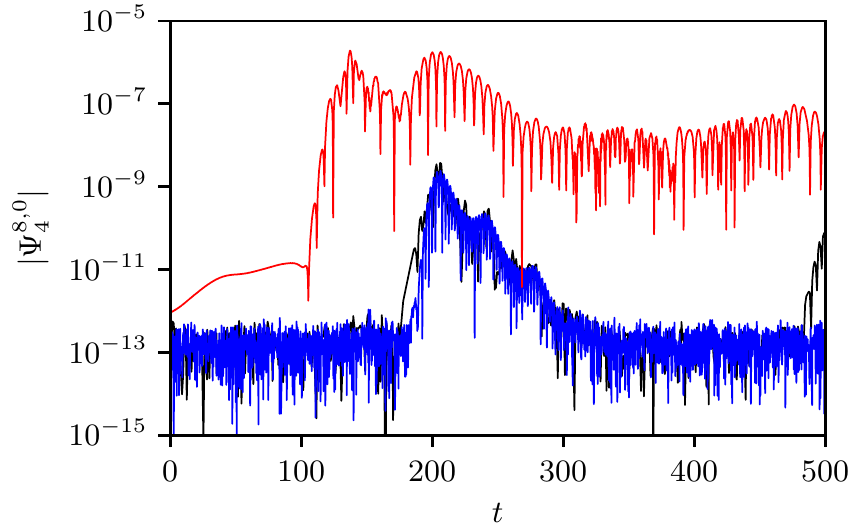}
	\end{tabular}
	\caption{$|\Psi_4|$ for the even $l=2$ through 8, $m=0$ modes,
	computed with both {\sc SphericalBSSN} (black and blue lines) and {\sc McLachlan} (red lines).}
	\label{fig:psi4_comparison}
\end{figure*}

In Fig.~\ref{fig:psi4_comparison}, we plot the absolute value of the multipole
expansion in spin-weighted spherical harmonics $_{-2}Y_{lm}$ of the Weyl scalar $\Psi_4$,
extracted at a radius of $r=180$, for the even $l=2$ through 8, $m=0$ multipoles
(by symmetry, only the $m=0$ modes are non-zero). The plot shows the 
Cartesian simulation with $dx=0.02$ and two different $\theta$-resolutions for the 
spherical simulations. There are notable differences between the Cartesian and spherical
evolutions: In the spherical simulations, there is an absence of
initial noise pulse
before the radiation reaches it peak value, and the decay after the peak value proceeds
much cleaner and to orders of magnitude below the values attained in the Cartesian 
simulation. The reason for this difference in behavior is the
fact that there are partial reflections of the outgoing wave at each Cartesian mesh 
refinement boundary (see, e.g.~\cite{Zlochower:2012fk,Etienne:2014tia}), causing the 
unphysical excitation of $l\geq4$ multipole modes, as well as reflections in the initial 
noise pulse in seen in the $l=2$ mode.  These reflections affect strong-field quantities 
as well, as described in~\cite{Etienne:2014tia}. Contrary to this, the
spherical grid in {\sc SphericalBSSN} is a single uniform grid, so there is a complete 
absence of these reflections (apart from reflections from the outer boundary), leading to much 
cleaner signals especially in the higher-order multipoles.  

\begin{figure*}
	\centering
	\begin{tabular}{p{.00\textwidth} >{\centering\arraybackslash}
         m{.5\textwidth} >{\centering\arraybackslash}
         m{.5\textwidth} }
	&\IncG[scale=1.0]{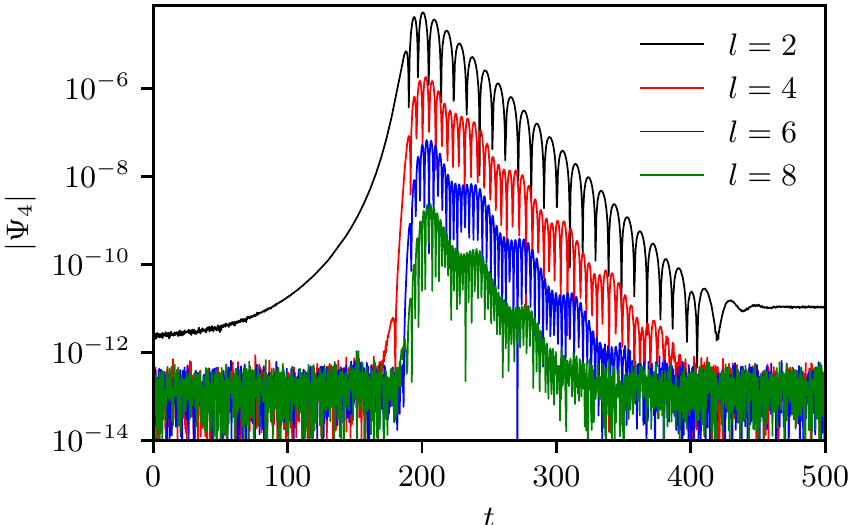}
	&\IncG[scale=1.0]{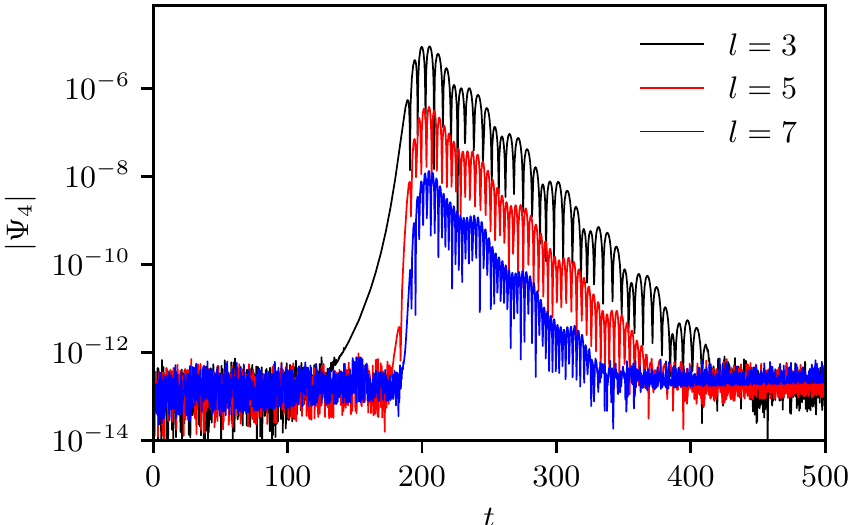}
	\end{tabular}
	\caption{$|\Psi_4|$ for the even $l=2$ through 8, $m=0$ modes (left panel), and
	the odd $l=3$ through 7, $m=0$ modes (right panel), computed with {\sc SphericalBSSN}.}
	\label{fig:psi4_sbssn}
\end{figure*}

\subsection{Kerr quasinormal modes}
\label{sec:QNM}
\begin{figure*}
	\centering
	\begin{tabular}{p{.00\textwidth} >{\centering\arraybackslash}
         m{.5\textwidth} >{\centering\arraybackslash}
         m{.5\textwidth} }
	&\IncG[scale=1.0]{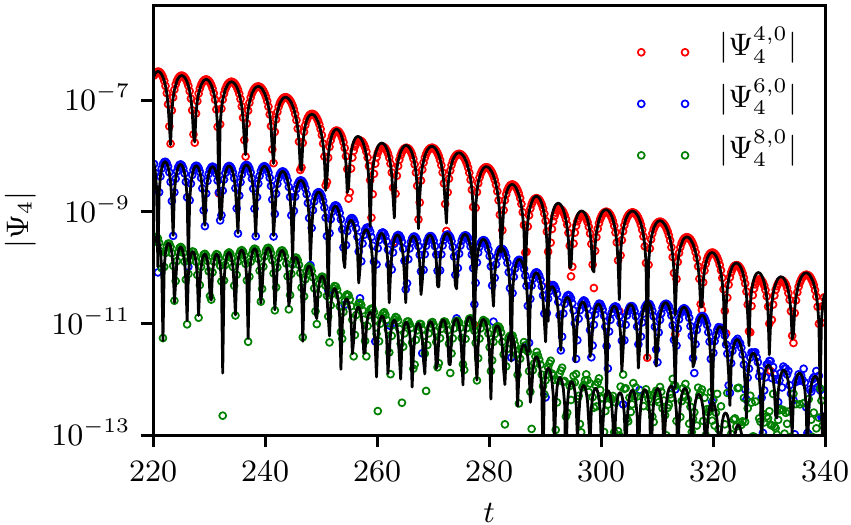}
	&\IncG[scale=1.0]{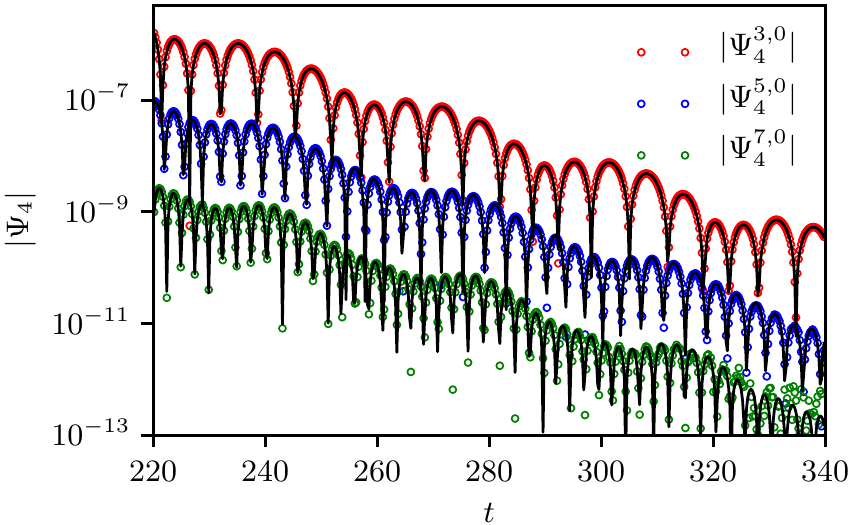}
	\end{tabular}
	\caption{Fits for the QNM ringdown, with even modes $l=2$ through $8$ in the left panel, 
	and odd modes $l=3$ through 7 in the right panel.  The solid lines represent the numerical 
	results for $|\Psi_4|$ as as computed with {\sc SphericalBSSN}, while the points are a 
	result of a fit (\ref{eq:qnm_fit}) to these numerical data.}
	\label{fig:psi4_qnm_fit}
\end{figure*}
Given that the {\sc SphericalBSSN} simulation provides us with very accurate
higher order modes, we turn our attention to an analysis of the BH's
quasinormal modes. As explained in the setup of the initial data, the spinning,
initially conformally flat BH should settle down to a Kerr BH via
ringdown of its quasinormal modes (QNM) (see~\cite{Berti2009} for a review).
In Fig.~\ref{fig:psi4_sbssn}, we plot the $l=2$ through 8, $m=0$ modes of $|\Psi_4|$
for the high resolution $N_{\theta}=64$ simulation alone.
We note that even (odd) $l$-modes contain only the real (imaginary) part
of $\Psi_4$. The $l=2$ mode follows a clear
exponential decay, but the higher-order modes exhibit a beating 
modulation on top of the exponential decay.  The reason for this behavior is
that the quasinormal modes for Kerr are defined in terms of spheroidal
harmonics, $_s{S}_{lm}$, while we decompose the waveform in terms of 
spin-weighted spherical harmonics $_sY_{lm}$.
Following~\cite{Teukolsky1972}, we can decompose $\Psi_4$ in terms of spin-weighted 
spheroidal harmonics $_sS_{l'm}$ with $s = -2$ according to
\begin{equation} \label{psi_exp_1}
\Psi_4 = \sum_{l',m',n'} A_{l'm'n'} \,{}_sS_{l'm'} e^{-\alpha_{l'm'n'} t} e^{i \omega_{l'm'n'} t} 
\end{equation} 
where $n$ is the overtone number of each mode, $\alpha_{l'mn}$ is its the decay rate, 
$\omega_{l'mn}$ its frequency, and the coefficients $A_{l'mn}$ are the amplitudes of 
the individual modes.  
In particular, we see that each mode oscillates and decays at well-defined rates.  
In practice, however, $\Psi_4$ is projected into the spin-weighted spherical 
harmonics ${}_sY_{lm}$, i.e.
\begin{eqnarray} \label{psi_exp_3}
\Psi_4^{lm} & = & \int \Psi_4 {}_s Y^*_{lm} d\Omega \nonumber \\
& = & \sum_{l',m',n'} A_{l'm'n'} e^{-\alpha_{l'm'n'} t} e^{i \omega_{l'm'n'} t} \int {}_sS_{l'm'} {}_sY^*_{lm} d\Omega \nonumber \\
& = & \sum_{l',m',n'} A_{l'm'n'} e^{-\alpha_{l'm'n'} t} e^{i \omega_{l'm'n'} t} \mu^*_{mll'n'}(a) \delta_{mm'} \nonumber \\
& = & \sum_{l',n'} A_{l'mn'} e^{-\alpha_{l'mn'} t} e^{i \omega_{l'mn'} t} \mu^*_{mll'n'}(a).
\end{eqnarray}
Here the coefficients $\mu^*_{mll'n'}(a)$ describe the mixing between spin-weighted
spheroidal and spherical harmonics; they are defined in eq.~(5) of~\cite{Berti2014} and depend on 
the black hole Kerr parameter $a$.

Ignoring higher-order overtone modes with $n' > 0$, which decay faster than 
the fundamental modes, we fit the $l=3$ through 8, $m=0$ spherical harmonic modes 
computed from our numerical data to the form
\begin{equation}\label{eq:qnm_fit}
\Psi^{l0}_4(t) \approx ~\sum_{l^{\prime}=2}^{l^{\prime}=10} 
A_{l^{\prime}0} e^{(-\alpha_{l^{\prime}00} t)}  {\rm sin} (\omega_{l^{\prime}00} t + \phi_{l^{\prime}0}),
\end{equation}
where the unknowns $A_{l^{\prime}0}$ and
$\phi_{l^{\prime}0}$ serve as 18 parameters corresponding to the amplitudes 
(including the mixing coefficients) and phases, respectively.
We fix the $\alpha_{l^{\prime}00}$ and $\omega_{l^{\prime}00}$ in the fit to be the
values of the decay rate and frequency that correspond to the Kerr BH in our
simulation ($J/M_{\rm BH}^2=0.573$ and $M_{\rm BH}=1.18112$), using the tabulated values 
and Mathematica notebooks to calculate QNMs~\cite{Berti2006,Berti2009} found at~\cite{BertiQNM}. 
The results of the fit for the $l=3$ through 8 modes are shown in 
Fig.~\ref{fig:psi4_qnm_fit} for a fitting window of $t=220-340$. 
The beating of the modes is very well captured by modeling a given $l$ mode as the sum 
of the expected decay rates and frequencies calculated in spin-weighted spheroidal harmonics, 
showing that mode mixing is responsible for the observed beating. A similar type of 
equal $m$ mode mixing has been observed in~\cite{Kelly2013}.

\section{Discussion}
\label{sec:conclusions}
We report on an implementation of the BSSN equations in spherical coordinates in 
the \ET.  While Cartesian coordinates have advantages for many applications, 
spherical coordinates are much better suited to take advantage of the approximate 
symmetries in many astrophysical systems. The problems associated with the coordinate 
singularities that appear in curvilinear coordinates can be avoided if these singularities 
are treated analytically -- which, in turn, is possible with the help of a reference-metric 
formulation of the BSSN equations~\cite{Bonazzola2004,Shibata2004,Brown2009b,Gourgoulhon2012,Montero2012b} 
and a proper rescaling of all tensorial quantities~\cite{Montero2012b,Baumgarte2013,Baumgarte2015,Ruchlin:2018com}. 
We implement this formalism in the \ET in an effort to make these techniques publicly 
available to the entire numerical relativity community and beyond.

Specifically, we adapt the \ET infrastructure, which originally was designed 
for Cartesian coordinates, for spherical coordinates. In contrast to Cartesian 
coordinates, spherical coordinates feature inner boundary condition, where ghost 
zones are filled by copying interior data from other parts of the numerical grid, 
taking into account proper parity conditions. We implemented these boundary conditions, 
which may require communication across processors, within an MPI-parallelized 
infrastructure using the {\sc Slab} thorn. Numerical code for the BSSN equations in 
spherical coordinates were provided by \SENR~\cite{Ruchlin:2018com,SENRNRPy:web}.

In order to test and calibrate our implementation we performed simulations of a single, 
spinning and initially conformally flat BH, and compared the evolution of BH mass, 
spin and GWs using our spherical BSSN ({\sc SphericalBSSN}) and Cartesian AMR BSSN 
({\sc McLachlan}) code with comparable grid resolutions.
For sufficiently high resolutions, the evolution on a spherical mesh conserves
irreducible mass and angular momentum far better than with Cartesian
AMR. In particular, there are no reflections
of the initial junk radiation or outgoing initial gauge
pulse~\cite{Etienne:2014tia,Zlochower:2012fk} at mesh refinement boundaries, causing
the evolution of irreducible mass and spin to be smoother in the unigrid spherical 
evolution. 
The advantage of using unigrid spherical coordinates over Cartesian coordinates with
box-in-box mesh refinement becomes particularly apparent when
analyzing the higher-order $l$-multipoles of the GW wave signal. These
signals are affected by partial reflections at mesh refinement 
boundaries, leading to a contamination of all higher order $l,m$ modes that never fully
leaves the computational domain. This effect is completely absent in the simulations using
the {\sc SphericalBSSN} thorn, where the quasinormal ringdown of the Kerr BH is 
observed to much smaller amplitudes than in the Cartesian simulations. We observe a 
significant beating of the exponential ringdown of multipoles with $l > 2$,
which can be explained by spheroidal-spherical multipole mode mixing. The accurate modeling
of the ringdown of higher order modes is necessary in order to provide
GW detectors with accurate templates~\cite{Berti2007}, as the
measurement of two quasinormal modes is needed to test 
the~\enquote{no hair theorem}~\cite{Dreyer2004,Berti2006}.

The current {\sc SphericalBSSN} thorn adopts uniform resolution in radius, which requires
a large number of points in order to place the outer boundary sufficiently far away and
to avoid contaminating the inner parts of the computational domain with noise 
from the outer boundary (i.e., to causally disconnect these inner parts from the outer 
boundary). Possible approaches to improve this is to adopt a non-uniform radial grid, 
e.g.~a logarithmic grid as implemented in~\cite{Baumgarte:2015aza} 
or~\cite{Sanchis-Gual:2015sxa}, or to use more general 
radial coordinates. The \SENR code~\cite{Ruchlin:2018com,SENRNRPy:web} allows for 
such generalized radial coordinates -- a convenient choice is $\sinh(r)$ -- and 
we plan to port these features into the {\sc SphericalBSSN} thorn in the
future.

We also plan to supplement our current implementation of Einstein's vacuum equations 
in spherical coordinates with methods for relativistic hydrodynamics and magnetohydrodynamics
as another set of publicly available thorns for the \ET.  
As shown in~\cite{Montero2014,Baumgarte2015}, these equations can be expressed with the 
help of a reference-metric as well.  Further using a rescaling of all tensorial quantities 
similar to the rescaling of the gravitational field quantities in this paper, the evolution 
of hydrodynamical variables is unaffected by the coordinate singularities.  We hope that 
with these methods, and possibly implementations of microphysical processes like radiation 
transport and nuclear reaction chains, the \ET in spherical coordinates will become a 
powerful and efficient community tool for fully relativistic simulations of a number of 
different objects, including rotating neutron stars, gravitational collapse, accretion 
disks, and supernova explosions.   
We believe that this will result in a new open source state-of-the-art code 
that will prove to be a valuable resource for a broad range of future simulations.

\begin{acknowledgments}
  The authors would like to thank Emanuele Berti for useful discussions and
  Dennis B. Bowen for a careful reading of the manuscript.
  We gratefully acknowledge the National Science Foundation (NSF) for
  financial support from Grants No.\  OAC-1550436, AST-1516150, PHY-1607520, PHY-1305730, 
  PHY-1707946,  PHY-1726215, to RIT, as well as Grants No.\
  PHYS-1402780 and PHYS-1707526 to Bowdoin College.
  V.M. also acknowledges partial support from AYA2015-66899-C2-1-P, and RIT for the 
  FGWA SIRA initiative.  
  This work used the Extreme Science and Engineering Discovery Environment (XSEDE) 
  [allocation TG-PHY060027N], which is supported by NSF grant No. ACI-1548562, and by the 
  BlueSky Cluster at RIT, which is supported by NSF grants AST-1028087, PHY-0722703, and PHY-1229173.
  Funding for  computer equipment to support the development of \SENR
  was provided in part by NSF EPSCoR Grant OIA-1458952 to West Virginia University.
  Computational resources were also provided by the Blue Waters sustained-petascale computing 
  NSF project OAC-1516125. 
\end{acknowledgments}

\bibliographystyle{apsrev4-1} \bibliography{references}

\end{document}